\documentclass[aps,preprint,showpacs,showkeys,nofootinbib,superscriptaddress]{revtex4}

\usepackage{graphicx}  %
\usepackage{amsmath}
\usepackage{bm}  %
\usepackage{dsfont}
\usepackage{ulem}

\newcommand{\simge}{\hspace*{0.2em}\raisebox{0.5ex}{$>$}
     \hspace{-0.8em}\raisebox{-0.3em}{$\sim$}\hspace*{0.2em}}

\newcommand{\bea}{\begin{eqnarray}}
\newcommand{\eea}{\end{eqnarray}}
\newcommand{\beq}{\begin{equation}}
\newcommand{\eeq}{\end{equation}}
\newcommand{\bqa}{\begin{eqnarray}}
\newcommand{\eqa}{\end{eqnarray}}

\def\mqo2{{\!\!\!}}

\begin{document}

\title{
Renormalization in the Three-body Problem\\ 
with Resonant P-wave Interactions}
\author{Eric Braaten}
\affiliation{Department of Physics,
         The Ohio State University, Columbus, OH\ 43210, USA\\}
\author{P. Hagen}
\author{H.-W. Hammer}
\affiliation{Helmholtz-Institut f\"ur Strahlen- und Kernphysik
	and Bethe Center for Theoretical Physics,
	Universit\"at Bonn, 53115 Bonn, Germany\\}
\author{L. Platter}
\affiliation{ Department of Fundamental Physics, Chalmers University of Technology, 
SE-412 96 Gothenburg, Sweden\\}
\date{\today}

\begin{abstract}
  Resonant P-wave interactions can be described by a minimal
  zero-range model defined by a truncated effective range expansion,
  so that the only 2-body interaction parameters are the inverse
  scattering volume $1/a_P$ and the P-wave effective range $r_P$.
  This minimal model can be formulated as a local quantum field theory
  with a P-wave interaction between atom fields and a molecular field.
  In the two-atom sector, the model is renormalizable, but it has
  unphysical behavior at high energies, because there are
  negative-probability states with momentum scale $r_P$.  In the
  sector with three atoms, two of which are identical, renormalization
  in some parity and angular-momentum channels involves an ultraviolet
  limit cycle, indicating asymptotic discrete scale invariance.  The
  Efimov effect occurs in the unitary limit $a_P^{-1/3}, r_P \to 0$,
  but this limit is unphysical because there are low-energy states
  with negative probability.  The minimal model can be of physical
  relevance only at energies small compared to the energy scale set by
  $r_P$, where the effects of negative-probability states are
  suppressed.
\end{abstract}

\smallskip
\maketitle

\section{Introduction}

\noindent
Systems with an S-wave resonance near a threshold arise 
in many fields of physics, including atomic, condensed matter, 
nuclear, and high-energy physics. 
They are important, because they exhibit universal behavior that 
depends on the energy of the resonance, but is otherwise insensitive 
to the nature of the constituents and the form of the interactions 
between them, as long as they have short range.  
The universal aspects of the few-body problem have been 
studied in detail \cite{Braaten:2004rn}.
The universal properties depend on the S-wave scattering length $a$.
If the Efimov effect occurs in the three-body problem, 
the universal properties also depend on a three-body scale parameter 
upon which observables can only depend log-periodically.
In the many-body problem with identical fermions with two spin states,
the universal properties have also been studied thoroughly,
theoretically and also experimentally using ultracold atoms
\cite{Giorgini:2008,Ketterle:2008,Lee:2008fa,Chin:2010aa}.
If the Efimov effect occurs in the three-body problem, 
the many-body problem is more complicated
and may not even be well-defined.

Systems with a P-wave resonance near a threshold also arise 
in many fields of physics.
A prominent example in nuclear physics is the low-energy 
P-wave resonance in $n-^4$He scattering which corresponds to
the unstable $^5$He nucleus \cite{Bertulani:2002sz}.
The universal aspects of such systems are not as clear-cut 
as in the S-wave case.  They involve at least two parameters:
the P-wave scattering volume $a_P$ and the P-wave effective range $r_P$.
The many-body physics of fermions with two spin states
that interact through a P-wave resonance has been studied using 
mean-field approximations \cite{HD:2005,O:2004,GRA:2004,CY:2005,GR:2006}.
If the P-wave interactions are strongly resonant,
the many-body physics may reveal qualitatively new features that are 
not captured by mean-field approximations \cite{Levi:2007}.

Ultracold atoms provide a promising laboratory for the experimental study 
of both few-body and many-body systems of particles with a P-wave 
resonance near threshold.  The inverse scattering length $a_P$
can be controlled experimentally  and can be made arbitrarily large 
by tuning the magnetic field to a P-wave Feshbach resonance. 
The first such studies were for 
a P-wave Feshbach resonance between identical fermions 
using $^{40}$K atoms \cite{Regal:2003B}.
Fermionic $^6$Li atoms have been studied near a Feshbach resonance 
between the lowest two hyperfine spin states \cite{Zhang:2004A,Schunk:2005A}.
The binding energies of P-wave dimers 
and the atom-dimer and dimer-dimer inelastic collision rates 
have been measured \cite{Vale:0802,Ueda:0803}.
P-wave Feshbach resonances have also been observed in a mixture of 
$^{40}$K fermions and $^{87}$Rb bosons \cite{Ferlaino:2006}.
One obstacle to studying the dependence on $a_P$
is that P-wave Feshbach resonances in ultracold atoms are usually very narrow.  

In this paper, we study the renormalization in the three-body sector 
for a minimal field-theoretic model of a system with a P-wave threshold
resonance.
In Section~\ref{sec:Twobody}, we explain why the model can only be physically
relevant at energies small compared to the energy scale set by $r_P$.
In Section~\ref{sec:PwaveSTM}, we write down the integral equation 
that can be used to calculate the bound-state spectrum 
of triatomic molecules in the model.
In Section~\ref{sec:RGlimitcycle}, we identify the angular-momentum 
and parity channels of the three-atom sector
in which a three-body parameter is required by renormalization.
Finally, in Section~\ref{sec:Conclusion}, we discuss whether the 
renormalization properties of the minimal model can be relevant 
to any real physical systems.

\section{Two-body problem with a P-wave resonance}
\label{sec:Twobody}

\noindent

We begin by discussing P-wave scattering at low energy
$E =k^2/(2\mu)$, where $\mu$ is the reduced mass.
If the particles interact through a short-range potential,
the P-wave phase shift has an effective range expansion 
in powers of $k^2$ that has the form
\beq
k^3 \cot \delta_P(k) =
-1/a_P  + \mbox{$\frac12$} r_P k^2 + \ldots\,,
\label{kcotdelta}
\eeq
%
where $a_P$ is the scattering volume and $r_P$ is the P-wave effective range,
which have dimensions (length)$^3$ and 1/(length), respectively.
If the only important interactions are in the P-wave channel 
and if the effective range expansion is truncated after the second term,
the scattering amplitude reduces to
\beq
f_k(\theta) =
\frac{k^2 \cos \theta}{-1/a_P  + \mbox{$\frac12$} r_P k^2 - i k^3}\,.
\label{f-k,theta}
\eeq
%
An equivalent expression is obtained if the interactions are dominated by
a coupling to a P-wave Feshbach resonance:  
\beq
f_k(\theta) =
\frac{k^2 \cos \theta}{(\nu - k^2/2 \mu)/g^2 - i k^3}\,.
\label{f-k,theta:Fesh}
\eeq
The parameter $r_P = - 1/\mu g^2$, which is negative definite,
controls the strength of the coupling to the Feshbach resonance. 
The detuning of the resonance from the
threshold is controlled by the combination $1/a_P r_P = \mu \nu$.
We will consider P-wave interactions for which the scattering 
amplitude has the minimal form 
in Eqs.~(\ref{f-k,theta}) or (\ref{f-k,theta:Fesh}) 
with the two parameters $a_P$ and $r_P$.

A point in the parameter space that is of particular interest is 
$a_P^{-1/3} = r_P = 0$, which is called the {\it unitary limit}
because the unitarity bound for scattering 
is saturated at this point.  Since they provide 
no length scale, the interactions are scale invariant in the unitary limit.
It is easy to see that the unitary limit cannot be realized 
with interactions through a short-range potential.
If the interaction potential vanishes outside the range $R$,
there is an upper bound on $r_P$ called the {\it Wigner bound} \cite{HL:plb09}.
For $|a_P|^{1/3} \gg R$, the Wigner bound reduces to $r_P \le -2/R$.
If we try to take the zero-range limit $R \to 0$ 
to justify the truncation of the effective range expansion in
Eq.~(\ref{kcotdelta}), $r_P$ is driven to $- \infty$.
The constraint $r_P < -2/R$ can also be derived by demanding
that the probability for a bound state to be in the region $r>R$ 
is less than 1 \cite{Pricoupenko,JLPC07}.

The behavior in the zero-range limit is improved if the 
potential with range $R$ is supplemented by a van der Waals tail 
that falls off like $-C_6/r^6$.  In this case, the expansion 
in Eq.~(\ref{kcotdelta}) also includes a linear term $b_P k$ \cite{Gao:pra98}.  
In the zero-range limit $R \to 0$, 
the coefficients $b_P$ and $r_P$ are determined 
by the scattering volume $a_P$ and the van der Waals length
$\beta_6 = (2 \mu C_6/\hbar^2)^{1/4}$.
If we also take the limit $|a_P|^{1/3} \gg \beta_6$, $b_P \to 0$ 
and $r_P \to  -3.4/\beta_6$.
Thus the low-energy expansion of $f_k(\theta)$ reduces to 
Eq.~(\ref{f-k,theta}), but the scale of $r_P$ is set by $1/\beta_6$.
The truncation of the expansion is justified only at low energies
$|E| \ll 1/\mu \beta_6^2$.

Real atoms interact through a short-range potential with a 
van der Waals tail, but they can also have couplings to diatomic molecules.
The scattering volume $a_P$ can be 
controlled and made arbitrarily large 
by tuning the magnetic field to a P-wave Feshbach resonance
where one of the molecules crosses the two-atom threshold.
The conditions for $f_k(\theta)$ to be well-approximated by the 
effective range approximation in Eq.~(\ref{f-k,theta}) have been 
studied by Zhang, Naidon, and Ueda \cite{ZNU:1010}.

Nishida has pointed out that the assumption that the scattering amplitude 
has the simple form in Eq.~(\ref{f-k,theta}) or 
(\ref{f-k,theta:Fesh}) up to arbitrarily large momentum $k$
necessarily implies the existence of states with negative 
probability \cite{Nishida:2011gs}.
In the case of a short-range potential, the problem is related 
to the Wigner bound.  If the bound $r_P < -2/R$ is violated,
the probability for a bound state to be in the region $r>R$ exceeds 1 
and that excess probability must be cancelled by negative probability
from the region $r < R$ \cite{Nishida:2011gs}.

The problem of negative-probability states can also be seen directly 
from the expression for the scattering amplitude in Eq.~(\ref{f-k,theta}).
It can be expressed as a function of the energy $E=k^2/2\mu$:
\beq
f_k(\theta) =
\frac{2 \mu E \cos \theta}
{-1/a_P  + \mu r_P E - (-2 \mu E - i \varepsilon)^{3/2}}\,.
\label{f-k,theta:E}
\eeq
%
The poles in $E$ of this function are the energies of bound states.
We set $1/a_P = 0$ for simplicity. If $r_P < 0$, as required by the Wigner bound,
the denominator then has zeroes at $E=0$ and $E=-r_P^2/8\mu$,
which correspond to one bound state at threshold 
and another with binding energy $r_P^2/8\mu$.
The scattering amplitude can be expressed as a sum
of contributions from the two poles and a function of $E$
that is regular at the poles:
\beq
f_k(\theta) = 2 E \cos \theta 
\left( \frac{1}{r_P E} - \frac{2}{r_P (E + r_P^2/8\mu)} 
+ ({\rm regular}) \right) \,.
\label{f-k,theta:poles}
\eeq
%
Note that the residues of the two poles have different signs.
The two bound states will contribute to unitarity sums
with the opposite sign of the residues in Eq.~(\ref{f-k,theta:poles}).
The probabilities of the bound states with binding energies 0 and 
$r_P^2/8\mu$ are positive and negative, respectively.
If $r_P>0$, there is only a negative probability pole with energy 0 on the physical sheet.
A negative-probability state with energy 0
is fatal for any physical interpretation of the threshold region.
If $r_P < 0$, the  negative-probability state with 
binding energy $r_P^2/8\mu$ is not necessarily fatal,
because its effects are suppressed at sufficiently low energy.
Thus a sensible physical interpretation requires $r_P < 0$
and $|E| \ll r_P^2/8\mu$.
If $1/a_P$ is nonzero, the analysis is more complicated. 
We refer to a pole as {\it shallow} if it has energy 
$|E| <  r_P^2/8\mu$ and as {\it deep} if it has energy $|E| \simge  r_P^2/8\mu$.
The different cases can be classified as:
\begin{enumerate}
\item $1/a_P < 0$: There is only one deep pole on the physical sheet. It 
has negative probability.
\item $0 < 1/a_P \leq |r_P|^3/54$: There are two poles on the physical sheet, 
one shallow pole with positive probability and one deep pole with 
negative probability.
\item  $1/a_P >|r_P|^3/54$: 
There are two poles with complex energies 
and complex residues on the physical sheet.
This case violates standard analyticity assumptions for the 
S-matrix and should be discarded  \cite{taylor}.
\end{enumerate}

The problem of atoms whose pair interactions give the scattering
amplitude in Eq.~(\ref{f-k,theta}) can be formulated as a quantum
field theory (for a review of field theoretical models of atom-atom
interactions see e.g. Ref.~\cite{Braaten:2007nq}) with two scalar fields $\psi_1$ and
$\psi_2$ (which annihilate atoms of types 1 and 2, respectively) and a
vector field $\bm{d}$ (which annihilates a diatomic molecule).  We
take the masses of the atoms to be $m_1$ and $m_2$.  The only
interaction is a P-wave contact interaction that allows transitions
between the diatomic molecule and a pair of atoms of types 1 and 2.
The form of the Lagrangian is constrained by Galilean invariance.  It
consists of kinetic terms, an energy offset for the diatomic molecule,
and a P-wave interaction term:
\bea
{\cal L} &=& 
\sum_{\sigma=1}^2 \psi_\sigma^\dagger
\left( i \frac{\partial~}{\partial t} 
+ \frac{1}{2 m_\sigma} \nabla^2 \right) \psi_\sigma
+ \eta \bm{d}^\dagger \cdot 
\left( i \frac{\partial~}{\partial t} 
+ \frac{1}{2 (m_1 + m_2)} \nabla^2 \right) \bm{d} 
\nonumber
\\
&& + \Delta_0 \bm{d}^\dagger \cdot \bm{d} 
- g_0 \mu \left[ \bm{d}^\dagger  \cdot 
(\psi_2 i\nabla \psi_1/m_1 - (i\nabla \psi_2) \psi_1/m_2) + ({\rm h.c.}) \right].
\label{Lag}
\eea
%
The parameters $\Delta_0$ and $g_0$ are bare parameters
that depend on the ultraviolet cutoff.
The parameter $\eta$ can be chosen by normalization of
the field $\bm{d}$ to be either $+1$ or $-1$.
If $\eta=+1$, $\bm{d}$ is an ordinary field which, in the absence of 
interactions, would annihilate an ordinary positive-probability molecule. 
If $\eta=-1$, $\bm{d}$ is a ghost field which, in the absence of 
interactions, would annihilate a negative-probability molecule.

The two-body problem for atoms 1 and 2 can be solved analytically.
The minimal model defined by the Lagrangian in Eq.~(\ref{Lag}) is 
renormalizable in the two-atom sector~\cite{Bertulani:2002sz}.
The bare parameters $g_0$ and $\Delta_0$ can be tuned as
functions of the ultraviolet momentum cutoff $\Lambda$
in such a way that the scattering amplitude reduces to
Eq.~(\ref{f-k,theta}) in the limit $\Lambda \to \infty$.
The renormalization conditions can be written
\begin{subequations}
\bea
\frac{1}{a_P} &=& 
\frac{6 \pi}{g_0^2 \mu} \Delta_0 + \frac{2}{3\pi} \Lambda^3,
\label{renorm:ap}
\\
\frac{r_P}{2} &=& 
- \eta \frac{3 \pi}{g_0^2 \mu^2} - \frac{2}{\pi} \Lambda.
\label{renorm:rp}
\eea
\label{renorm}
\end{subequations}
%
In order to have finite limits as $\Lambda \to \infty$,
the two terms on the right sides of both Eqs.~(\ref{renorm:ap})
and (\ref{renorm:rp}) must have opposite signs.
A finite limit for $a_P$ requires that $\Delta_0$ be negative 
and that $\Delta_0/g_0^2$ scale like $\Lambda^3$.
A finite limit for $r_P$ requires that $g_0^2$ scale like 
$\Lambda^{-1}$ and that $\eta = -1$. 
Thus $\bm{d}$ must be a ghost field.  If there were no interactions, 
it would annihilate a negative-probability molecule.
To see whether there are any 
negative-probability particles in the presence of interactions,
one needs to examine the scattering amplitude in Eq.~(\ref{f-k,theta}).
As we have seen, $r_P < 0$ is necessary to avoid 
negative-probability particles near the two-atom threshold.
The energy restriction $|E| \ll r_P^2/8 \mu$ is then necessary to ensure 
that unphysical effects associated with negative-probability particles
are suppressed.
The simplest observable that can go 
wrong is the two-body bound-state spectrum. 
It contains unphysical low-energy states if the 
energy restriction is not observed.

Nishida also presented an algebraic argument for the existence of states 
with negative probability in this model \cite{Nishida:2011gs}.
In the unitary limit $a_P^{-1/3} = r_P = 0$, 
the model has scaling symmetry as well as Galilean symmetry.
It therefore also has nonrelativistic conformal 
symmetry \cite{Nishida:2007pj}.
The asymptotic behavior of the scattering amplitude 
in Eq.~(\ref{f-k,theta}) at large $k$ implies that the dimer field
$\bm{d}$ has scaling dimension 1. 
However Nishida and Son have used the nonrelativistic conformal
algebra together with the assumption that all states have positive
norm to prove that primary operators can only have 
scaling dimensions greater than or equal to 3/2.
The violation of the bound in this model implies that 
there must be states with negative norm.

\section{P-wave STM Equation}
\label{sec:PwaveSTM}

\noindent

We now consider the three-body problem with resonant P-wave interactions
between pairs of atoms.
For three identical bosons, there can be no P-wave interactions.
The case of  three identical fermions has been studied thoroughly by 
Jona-Lasinio, Pricoupenko, and Castin \cite{JLPC07}.
They considered a two-channel model with atoms and a diatomic molecule 
that have interactions with a finite range $b$.
They found that $r_P$ has to be negative and that there is a lower
bound on $|r_P|$ that is proportional to $1/b$.
In the three-fermion sector, they calculated the spectrum of triatomic molecules
(trimers), atom-dimer scattering lengths, and three-body recombination rates.
For $|a_P| \gg b^3$, the trimer spectrum at energies small compared to 
$\hbar^2/mb^2$ can consist of either 0 or 1 trimer with positive parity 
and either 0 or 1 trimer with negative parity.
Three-body recombination was also studied previously by Suno, Esry, and Greene
using the adiabatic hyperspherical representation of the Schr\"odinger equation
for three identical fermions interacting thorough a short-range potential 
that is tuned to give a large P-wave scattering volume \cite{SEG:2002}.

In the next simplest case of a three-body problem 
with resonant P-wave interactions, there are
two identical atoms of type 2 and a third atom of type 1.
Atoms of types 1 and 2 interact through a P-wave resonance
and we assume that the interaction between the identical atoms 
can be neglected.
We describe this system using the model defined by the Lagrangian 
in Eq.~(\ref{Lag}). 
Transition amplitudes for three atoms
can be calculated exactly numerically by solving a
single-variable integral equation that is analogous to the 
Skornyakov--ter-Martyrosian
(STM) equation for S-wave contact interactions~\cite{STM57}.
The equations for different orbital angular momentum and parity
quantum numbers $J^P$ can be decoupled. 
The bound-state problem reduces to solving an eigenvalue equation 
for the energy $E<0$ that is a homogeneous integral equation of the form
\beq
B(p) = \nu \frac{3(-1)^{J+1}}{\pi(2J+1)}
\int_0^\Lambda \text{d}q \, R^{J^P}(p,q,E) D(q,E) B(q),
\label{fB-inteq}
\eeq
%
where the prefactor $\nu$ is $+1$ or $-1$ if the identical atoms 
are bosons or fermions, respectively. 
The upper limit $\Lambda$ of the integral is an ultraviolet 
momentum cutoff that should be taken to $\infty$.
The dimer propagator $D(q,E)$ is given by
\begin{subequations}
\bqa
D(q,E) &=& q^2 
\left[ \frac{1}{a_P} + \frac{r_P}{2} b(q,E) + b(q,E)^{3/2} \right]^{-1},
\\
b(q,E) &=& \frac{r(2+r)}{(1+r)^2} \left[q^2 - \frac{2(1+r)}{2+r}m_2E\right],
\eqa
\end{subequations}
%
where $r = m_1/m_2$ is the mass ratio.
The kernel $R^{J^P}(p,q,E)$ depends on the quantum numbers $J^P$.
For positive parity $J^+$ with $J>0$, 
the bound-state amplitude $B(p)$ is a two-component 
column vector and $R^{J^+}(p,q,E)$ is a $2 \times 2$ matrix with entries
\begin{subequations}
\bea
R^{J^+}_{11} &=& \left[\frac{J}{1+r} +\frac{1+r}{2J-1}\right] Q_J 
 - J\left[\frac{p}{q} + \frac{q}{p} \right]Q_{J-1} 
\nonumber \\ 
&& + \frac{(J-1)(2J+1)(1+r)}{2J-1}Q_{J-2},
\\
R^{J^+}_{12} &=& 
\sqrt{J(J+1)} \big( -[1/(1+r) + 1+r] Q_J 
\nonumber \\ 
&&
+ (q/p) Q_{J-1} + (p/q) Q_{J+1} \big),
\\
R^{J^+}_{22} &=& \left[\frac{J+1}{1+r} +\frac{1+r}{2J+3}\right] Q_J 
 - (J+1) \left[\frac{p}{q} + \frac{q}{p} \right] Q_{J+1}
\nonumber \\
&& 
+ \frac{(J+2)(2J+1)(1+r)}{2J+3}Q_{J+2},
\label{R-matrix:J+}
\eea
\end{subequations}
%
and $R^{J^+}_{21}(p,q,E) = R^{J^+}_{12}(q,p,E)$.
For $J \ge 0$, $Q_J(x)$ is a Legendre function 
of the second kind and its argument is $x = (1+r)[p/q+q/p]/2-m_2 r E/(pq)$. 
For $J < 0$, $Q_J = 0$.
For the special case $J^P=0^+$ and for negative parity $J^-$ 
with $J>0$, $B(p)$ has one component
and $R^{J^P}$ is a $1 \times 1$ matrix.
For $0^+$, its entry 
$R^{0^+}_{22}(p,q,E)$ is given by Eq.~(\ref{R-matrix:J+}). 
For $J^-$, its entry is
\beq
R^{J^-}(p,q,E) = (1+r) \left[ Q_{J+1}(x) - Q_{J-1}(x) \right].
\label{M-matrix:J-}
\eeq
These equations are valid for $1/a_P \leq |r_P|^3/54$.

\section{Renormalization group limit cycles}
\label{sec:RGlimitcycle}

\noindent

In most $J^P$ channels, the solution to the bound state equation
for $B(p)$ decreases rapidly enough at large $p$ that the integral 
in Eq.~(\ref{fB-inteq}) converges as $\Lambda \to \infty$.
In these channels, the model is renormalizable:
the only parameters that are required are $a_P$ and $r_P$.
However there are also channels in which $B(p)$ approaches a 
log-periodic function of $p$ at large $p$, just like in the STM equation.  
In this case, the integral does not converge, but instead approaches 
a log-periodic function of $\Lambda$ as $\Lambda \to \infty$.
If the identical particles are bosons, 
the only channel in which this behavior occurs is $1^+$.
If the identical particles are fermions, 
this behavior occurs in the channels $0^+$, $1^+$, $1^-$ and $2^+$.

In the field theory framework, the cutoff dependence implies that the 
model defined by the Lagrangian in Eq.~(\ref{Lag}) is not renormalizable 
in the three-atom sector.  In the case of an S-wave contact interaction,
renormalizability can be 
restored by adding a three-body contact interaction 
term \cite{Bedaque:1998kg}.  
In our case with a P-wave interaction, renormalizability can 
also be restored by adding an appropriate 
three-body contact interaction for each $J^P$
channel in which there is log-periodic dependence on $\Lambda$.
The renormalization group (RG) flow for its coupling constant 
is governed by an RG limit cycle \cite{Braaten:2004rn}.
An RG limit cycle is characterized by a discrete scaling factor $\lambda_0$
such that every time $\Lambda$ increases by a factor of $\lambda_0$,
the coupling constant returns to its original value.

Renormalization can be implemented by eliminating the coupling constant 
for the three-atom interaction in favor of a renormalization scale 
$\Lambda_*$ \cite{Bedaque:1998kg}.
The dependence of any renormalized amplitude on $\Lambda_*$
can only be log-periodic with discrete scaling factor $\lambda_0$.
The model therefore exhibits asymptotic discrete scale invariance
at high energies.
A discrete scale transformation consists of rescaling all momenta 
and energies by factors of $\lambda_0^n$ and $\lambda_0^{2n}$, 
respectively, and also rescaling the parameters
as $a_P \to \lambda_0^{-3n} a_P$ and $r_P \to \lambda_0^n r_P$,
where $n$ is an integer.
Under a discrete scale transformation, any amplitude changes 
simply by an overall power of $\lambda_0^n$.

The most dramatic consequence of the discrete scale invariance
is the Efimov effect.
Efimov showed in 1970 that if the two-body S-wave scattering length 
is tuned to infinity, there can be an infinite sequence of three-body 
bound states called {\it Efimov states}
that have an accumulation point at the three-body threshold~\cite{Efimov70}.
The binding energies of successive 
states differ by the square of a 
discrete scaling factor $\lambda_0$ that depends on the mass ratios 
and symmetries.  The cases in which the Efimov effect occurs include 
three identical bosons, 
two identical bosons  and a third particle, 
and two identical fermions and a third particle whose mass is smaller 
than the fermion mass divided by 13.6 (see, e.g.~\cite{Braaten:2004rn}).
In the case of resonant P-wave interactions between one atom and 
two identical atoms, we have identified all the channels 
in which the Efimov effect occurs.
If the identical particles are bosons, 
the Efimov effect occurs only in the $1^+$ channel.
If the identical particles are fermions, 
the Efimov effect  occurs only 
in the channels $0^+$, $1^+$, $1^-$ and $2^+$.
The Efimov effect in the $1^-$ sector was discovered 
by Macek and Sternberg \cite{MS:prl06,macek:2007}. Our results for
the $1^-$ sector are consistent with their results.
Unfortunately the Efimov effect in this model is unphysical.
In the unitary limit, there are negative-probability states 
at the three-atom threshold, which is the accumulation point for the 
infinitely many Efimov states.

\noindent

The discrete scaling factor for each $J^P$ channel
can be expressed as $\lambda_0=\exp(\pi/s_0)$, 
where $i s_0$ is a pure imaginary solution to the equation
\beq
\det\left[\mathds{1}-\nu \, \frac{3}{\pi} \frac{(-1)^{J+1}}{2J+1} 
\left(\frac{(1+r)^2}{r(2+r)}\right)^{\frac32}\,M^{J^P}(s)\right] = 0
\label{det1_M}
\eeq
and $M^{J^P}$ is essentially the Mellin-transform of $R^{J^P}$ in 
Eq.~\eqref{fB-inteq}. For positive parity $J^+$ with $J>0$, 
$M^{J^+}(s)$ is a $2 \times 2$ matrix with entries
\begin{subequations}
\bea
M^{J^+}_{11} &=& \left[\frac{J}{1+r} +\frac{1+r}{2J-1}\right] f_J 
- J\left[\hat t^{-} + \hat t^+ \right]f_{J-1} 
\nonumber \\
&& 
+ \frac{(J-1)(2J+1)(1+r)}{2J-1} f_{J-2},
\\
M^{J^+}_{12} &=& \sqrt{J(J+1)} \big(  - [1/(1+r) + 1+r] f_J
\nonumber \\
&&
+ \hat t^+ f_{J-1} + \hat t^{-} f_{J+1} \big),
\\
M^{J^+}_{22} &=& \left[\frac{J+1}{1+r} +\frac{1+r}{2J+3}\right] f_J
- (J+1) \left[\hat t^{-} + \hat t^+ \right] f_{J+1}
\nonumber \\
&& 
+ \frac{(J+2)(2J+1)(1+r)}{2J+3}f_{J+2},
\label{M-matrix:J+}
\eea
\end{subequations}
%
and $M^{J^+}_{21} (s) = M^{J^+}_{12}(-s)$.
The functions $f_J(s)$ are 
\bea
f_J(s) &=& P_J\left(\frac{1+r}{2}[\hat t^{-} + \hat t^+]\right)f_0(s),
\eea
%
where $P_J$ is a Legendre polynomial if $J \ge 0$ and 0 if $J<0$, 
$\hat t^{\pm}$ is an operator
that shifts the argument of $f_J(s)$ by $\pm 1$, 
and $f_0(s)$ is 
\beq
 f_0(s) = \frac{\pi \sin(s \arcsin[1/(1+r)])}{s \cos(s \pi/2)}.
\label{f-s}
\eeq
%
In the special case $J^P=0^+$ and for negative parity $J^-$ with $J>0$,
$M^{J^P}(s)$ is a $1 \times 1$ matrix.
For $0^+$, its entry 
$M^{0^+}_{22}(s)$ is given by Eq.~(\ref{M-matrix:J+}). 
For $J^-$, its entry is
\beq
M^{J^-}(s) = (1+r) \left[ f_{J+1}(s) - f_{J-1}(s) \right].
\label{R-matrix:J-}
\eeq
%

\begin{figure}[t]
\centerline{\includegraphics*[width=10cm,angle=0,clip=true]{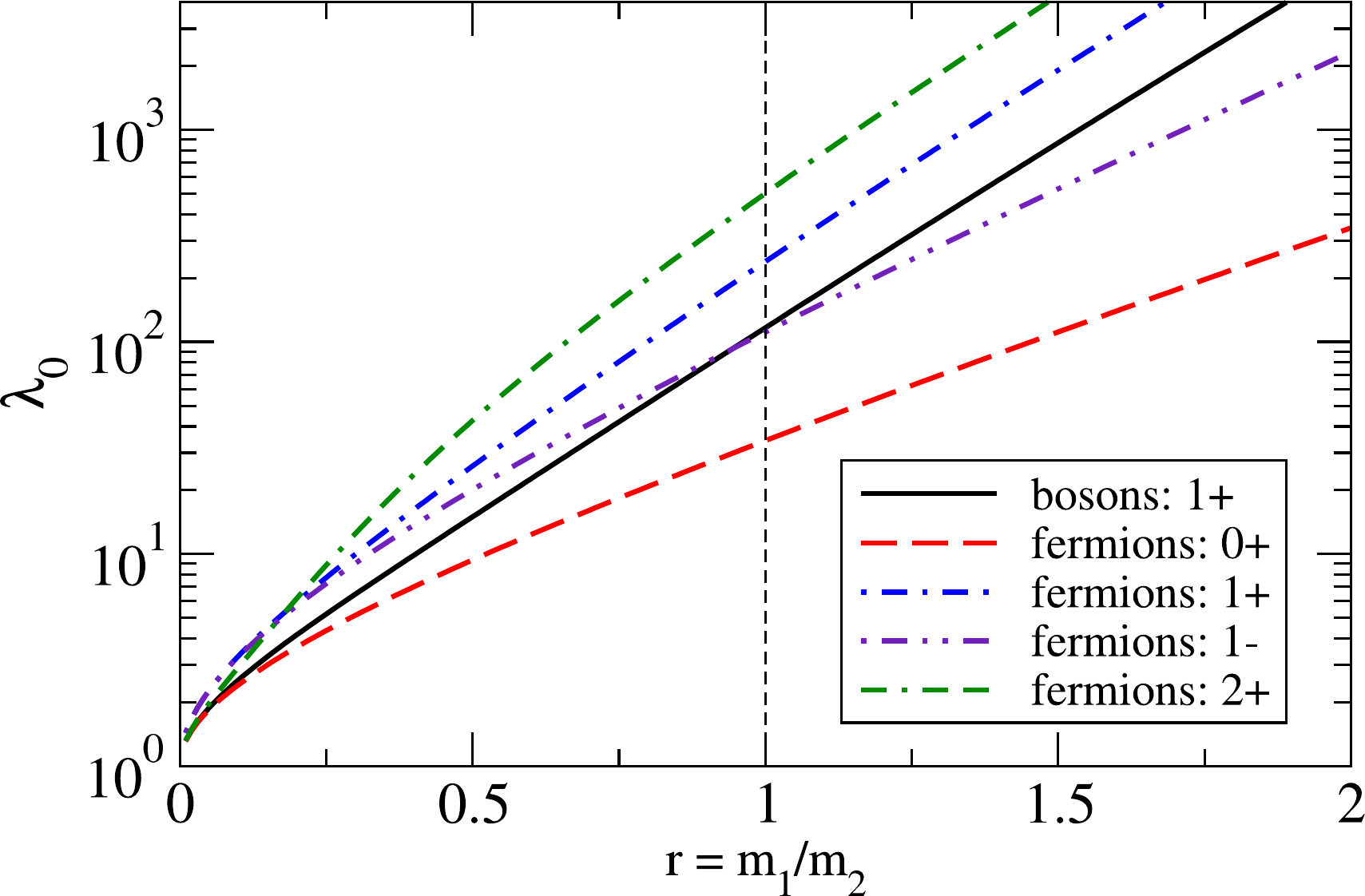}}
\vspace*{0.0cm}
\caption{(Color online)
Discrete scaling factor $\lambda_0 =\exp(\pi/s_0)$ as a function of 
the mass ratio $r=m_1/m_2$.
If the identical particles are bosons, 
the Efimov effect occurs only in the $J^P = 1^+$ sector.
If the identical particles are fermions, 
the Efimov effect occurs in the $0^+$, $1^+$,  $1^-$, and $2^+$ sectors.
}
\label{fig:dsf}
\end{figure}

If the identical atoms are bosons, Eq.~\eqref{det1_M} has a 
pure imaginary solution $is_0$ only for $J^P = 1^+$.
For equal masses ($r=1$),
the discrete scaling factor $\lambda_0= \exp(\pi/s_0)$ is 116.7.
If the identical atoms are fermions, Eq.~\eqref{det1_M} has a 
pure imaginary solution for $0^+$, $1^+$, $1^-$, and $2^+$.
For equal masses, the discrete scaling factors 
are 34.2, 239.3, 111.2, and 501.9, respectively.
The dependence of $\lambda_0$ on the mass ratio 
$r= m_1/m_2$ for each of the five cases is shown in Fig.~\ref{fig:dsf}.
The discrete scaling factors all approach 1 as $r \to 0$.

We proceed to discuss the spectrum of the three-body bound states 
at $r_P = 0$ predicted by Eq.~(\ref{fB-inteq}).
In the unitary limit defined by $a_P^{-1/3} = r_P = 0$,
discrete scale invariance implies the existence of infinitely many 
Efimov trimers.  They have an 
accumulation point at the three-atom threshold, and the binding energies of
successive trimers differ by a factor of $\lambda_0^2$.
As $|a_P|^{-1/3}$ increases, each of the Efimov trimers disappears 
through the three-atom threshold at a negative $a_P$
and through the atom-dimer threshold at a positive $a_P$.
Discrete scale invariance requires that if an Efimov trimer 
disappears through a threshold at $(a_P,r_P)=(a_P^*,r_P^*)$,
the next deeper Efimov trimer must disappear through the same 
threshold at $(\lambda_0^{-3}a_P^*,\lambda_0 r_P^*)$.

\begin{figure}[t]
\centerline{\includegraphics*[width=10cm,angle=0,clip=true]{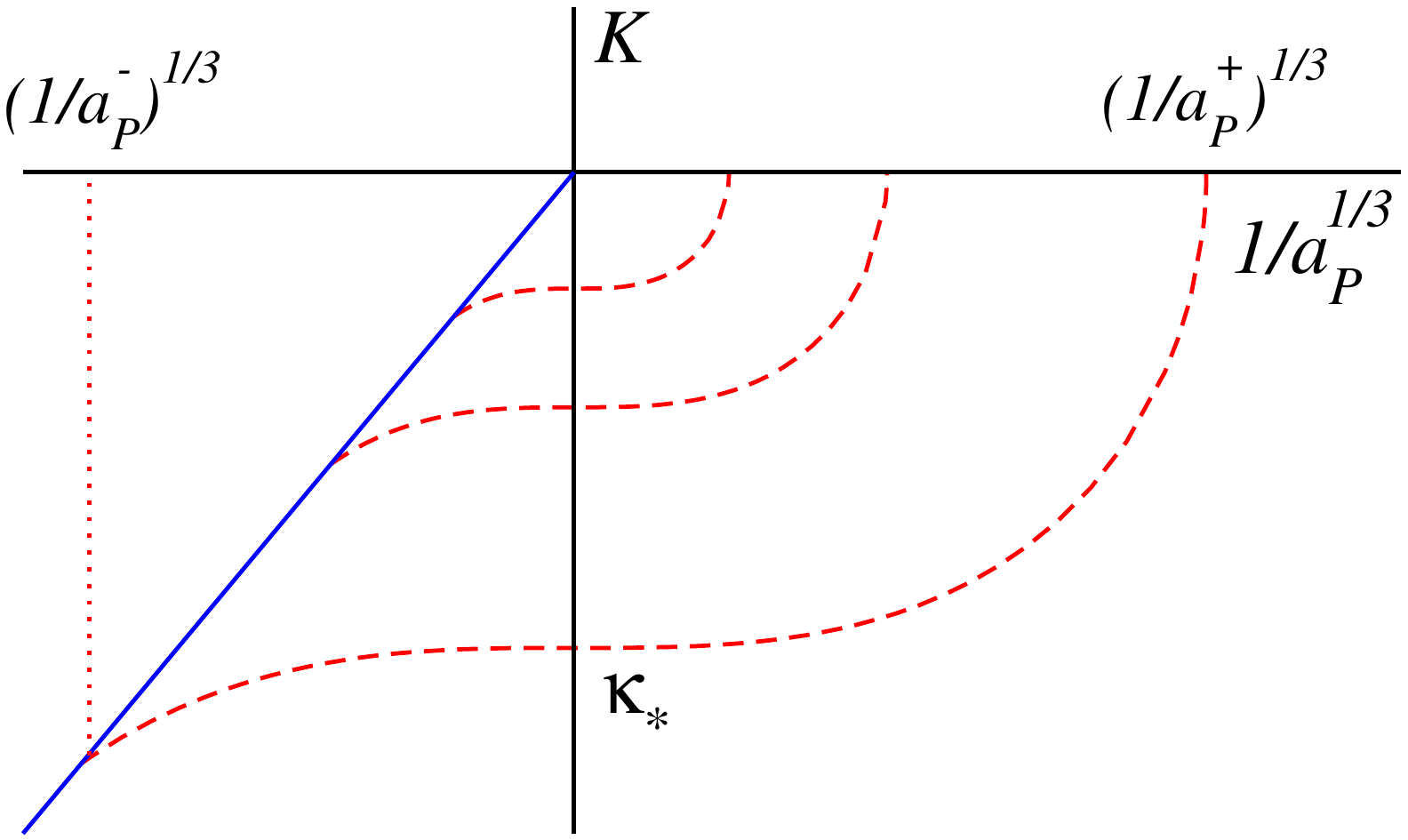}}
\vspace*{0.0cm}
\caption{(Color online)
Efimov plot showing the energy variable $K$ as a function of 
$(1/a_P)^{1/3}$ with $r_P=0$ for the case of 
identical fermions, equal masses, and $J^P=0^+$.
The scaling factor $\lambda_0\approx 34.2$ was divided by 17
to fit more states in the plot.
}
\label{fig:Eplot}
\end{figure}
The bound state spectrum at $r_P = 0$ predicted by Eq.~(\ref{fB-inteq})
is illustrated in Fig.~\ref{fig:Eplot} for the case with
identical fermions, equal masses, and quantum numbers $0^+$.
The energy variable $K ={\rm sign}(E) (\mu |E|/\hbar^2)^{1/2}$ 
for several Efimov trimers is shown as a 
function of $a_P^{-1/3}$ for $r_P=0$.
An Efimov trimer whose binding energy at unitarity is 
$\hbar^2\kappa_*^2/m$ disappears 
through the three-atom threshold at $a_P^- = -0.52/\kappa_*^3$ 
and through the atom-dimer threshold at $a_P^+ = +0.24/\kappa_*^3$.
The behavior in the other channels is similar.
The discrete scaling factors for equal masses and the 
corresponding threshold parameters for all channels $J^P$ are given in 
Table~\ref{tab:1sf}.
%
\begin{table}[t]
\begin{tabular}{cc|cccc}
identical particles & $J^P$ & $s_0$ & $\lambda_0$ 
& $a_P^- \kappa_*^3$ & $a_P^+ \kappa_*^3$
\\
\hline
bosons & ~$1^+$~ & ~$0.660$~ & ~$116.7$~ & ~$-0.134(2)$~ & ~$0.452(2)$~
\\
fermions & $0^+$ & $0.889$ & $~34.2$ & ~$-0.522(2)$~ & ~$0.244(2)$~
\\
fermions & $1^+$ & $0.574$ & $239.3$ & ~$-0.25(2)$ & ~$0.34(2)$~
\\
fermions & $1^-$ & $0.667$ & $111.2$ & ~$-0.576(2)$~ & ~$0.212(2)$~
\\
fermions & $2^+$ & $0.505$ & $501.9$ & ~$-0.590(1)$~ & ~$0.19(3)$~
\\
\end{tabular}
\caption{The imaginary solutions $is_0$ of Eq.~\eqref{det1_M} 
and the discrete scaling factors $\lambda_0$ 
for the case of equal masses
$m_1=m_2$. The threshold parameters $a_P^-$ and $a_P^+$
for $r_P =0$ are also given.
}
\label{tab:1sf}
\end{table}
%

\section{Conclusion}
\label{sec:Conclusion}

\noindent

Whether there is a universal binding mechanism for shallow P-wave
states remains an intriguing and important question. 
Apart from ultracold atoms,
such systems occur frequently in halo nuclei~\cite{Jensen:2004}.
The application of effective field theory to such a system is
called {\it halo EFT}~\cite{Bertulani:2002sz}.

In this paper, we have discussed the renormalization of the minimal
zero-range model for P-wave interactions defined by
Eq.~(\ref{Lag}). This model is renormalizable in the two-body sector,
but it has unphysical negative-norm molecules with momentum scale
$r_P$. It can be used as an effective theory for typical momenta well
below the scale $r_P$ where the effects of negative probability states
are suppressed.  In the three-body sector, renormalization in some
parity and angular-momentum channels involves an ultraviolet limit
cycle, indicating asymptotic discrete scale invariance for large
momenta.  In the unitary limit $a_P^{-1/3}, r_P \to 0$, there is an
accumulation of three-body states at threshold analog to the Efimov
effect. However, this limit is unphysical since the negative norm
states are driven to threshold and lead to the breakdown of the model.

\begin{figure}[t]
\centerline{\includegraphics*[width=9cm,clip=true]{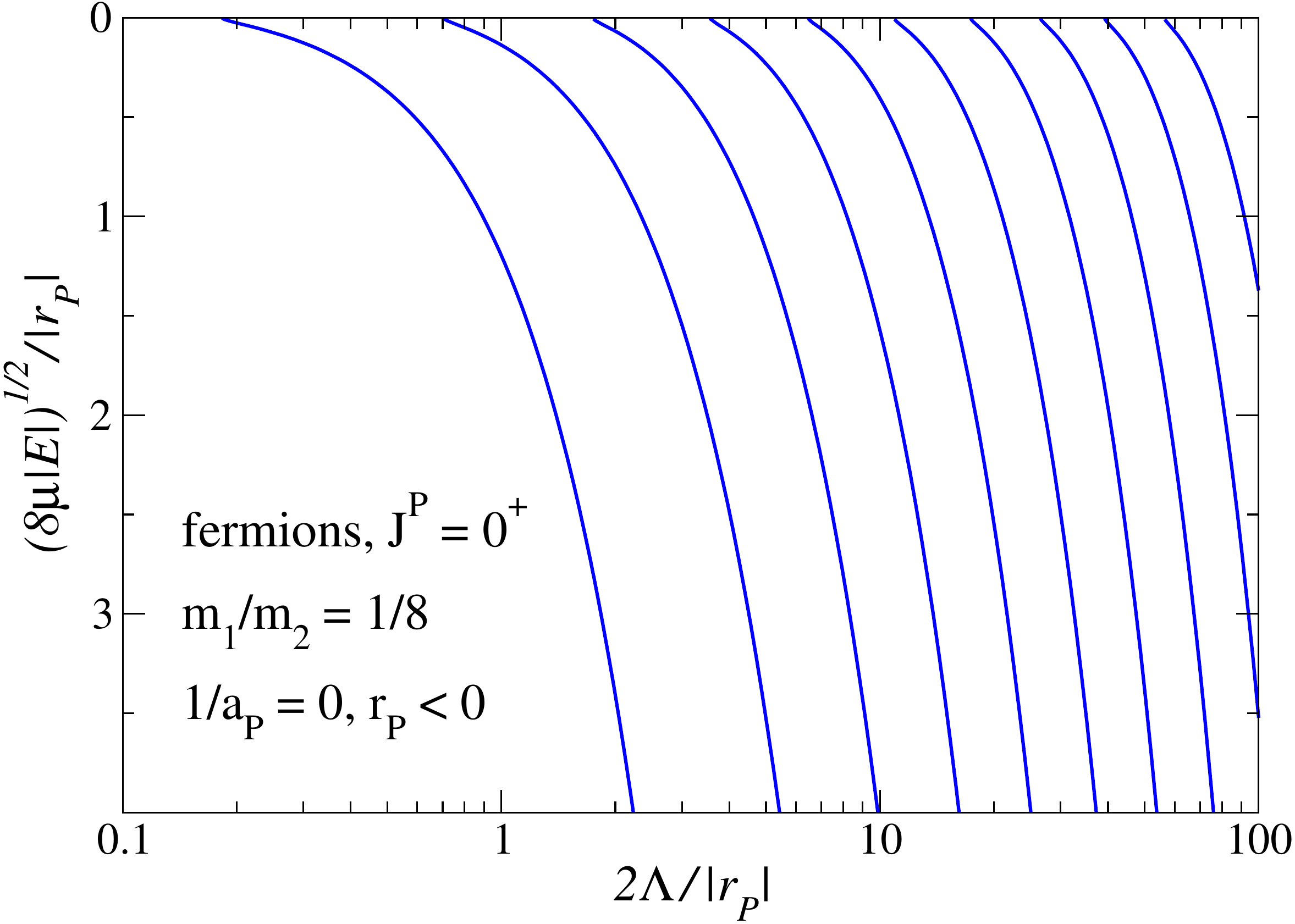}}
\vspace*{0.0cm}
\caption{(Color online)
The dimensionless binding momentum $\sqrt{8\mu|E|}/|r_P|$
as a function of the dimensionless cutoff $2\Lambda/|r_P|$ for fermions with 
$J^P = 0^+$, $1/a_P =0$, $r_P <0$, and $m_1/m_2=1/8$.
}
\label{fig:dimerexpand}
\end{figure}

A promising strategy for dealing with P-wave resonant interactions
is to treat the unitarity term $i k^3$ from Eq.~(\ref{f-k,theta})
as a perturbation~\cite{Bedaque:2003wa}. Such
an expansion can be justified rigorously for small momenta,
$k \ll |r_P|$,  and leads
to a simplified pole structure of the dimer propagator. 
As an example, we show in Fig.~\ref{fig:dimerexpand} the 
dimensionless binding momentum $\sqrt{8\mu|E|}/|r_P|$ as
a function of the dimensionless cutoff $2\Lambda/|r_P|$ for fermions
in the $0^+$-channel, with $1/a_P =0$, $r_P <0$, and $m_1/m_2=1/8$. 
There can be 0, 1, or possibly even 2 three-body bound states for
dimensionless cutoffs below 1 where the expansion of the dimer
propagator can be justified.  Increasing the mass ratio $m_1/m_2$ also
increases the critical value of the dimensionless cutoff at which 
the first three-body bound state appears.  For equal masses, this critical
value is larger than 1, so there are no such states without
introducing a three-body force.

The strategy of treating the unitarity term as a 
perturbation was recently applied to $^6$He by Ji, Elster, and
Phillips~\cite{Ji:2011}. This halo nucleus
consists of an alpha particle and two weakly-bound neutrons and can be
described as an effective three-body system. The alpha-neutron
interaction has a strong P-wave resonance while the two neutrons
interact resonantly in an S-wave channel. 
Proper renormalization of the halo EFT was achieved by introducing an
appropriate three-body force
that was tuned to give the measured binding energy of $^6$He~\cite{Ji:2011}. 
Rotureau and van Kolck~\cite{Jimmy:2011,Rotureau:2012yu} described 
$^6$He in the Gamow shell model using a different strategy. They 
applied a halo effective theory with separable neutron-alpha
and neutron-neutron interactions and a neutron-neutron-alpha
three-body force. The range of the separable interactions then acts as an 
ultraviolet cutoff. The unitarity term can be treated non-perturbatively
without introducing any negative-probability states. Similar 
strategies have been used for the pionless EFT in the three-nucleon system.

The successful description of the $^6$He nucleus within halo 
EFT~\cite{Ji:2011}
suggests that the field theory presented in this work
can describe three-body phenomena in experiments with ultracold atoms,
provided that the unitarity term in the P-wave scattering amplitude is
treated as a perturbation. Future applications might
include the calculation of the scattering-volume dependence of the
three-body recombination rate and other few-body reaction rates.

\begin{acknowledgments}
We thank Y.~Nishida for pointing out that the minimal zero-range model
for resonant P-wave interactions has negative-probability states.
We also thank D.~Phillips for valuable discussions.
This research was supported in part 
by a joint grant from the ARO and the AFOSR,
by the DFG through SFB/TR 16, 
by the BMBF under contract No.~06BN9006, 
and by the Swedish Research Council.
\end{acknowledgments}

\end{document}